\documentclass[showpacs,amsmath,amssymb]{revtex4}
\usepackage{amsmath}
\usepackage{graphicx}
\usepackage{dcolumn}
\usepackage{bm}

\begin{document}

\title{Landau Levels Analog  to Electric Dipole}
\author{L. R. Ribeiro, Claudio Furtado$^{1}$\footnote{Electronic Address:
furtado@fisica.ufpb.br} and J. R. Nascimento$^{2}$\footnote{Electronic Address:
jroberto@fisica.ufpb.br}}

\affiliation{Departamento de F\'{\i}sica, CCEN,  Universidade Federal
da Para\'{\i}ba, Cidade Universit\'{a}ria, 58051-970 Jo\~ao Pessoa, PB, Brazil}

\begin{abstract}
In this letter we study the quantum dyamics of a neutral particle in the presence of an external magnetic field. We demonstrate in a specific field-dipole configuration that we have a  quantization similar to the Landau Levels. We investigate this quantization motivated by the recent analysis of Landau-Aharonov-Casher(LAC) quantization of Ericsson and Sj\"oqvist[Phys Rev. A {\bf 65} 013607 (2001)]. The energy eigenfuction and eigenvalues are obtained. 
\end{abstract}


\maketitle
\newpage

The study of the quantum dynamics of charged and neutral particles in the presence of electromagnetic fields is responsible for a series of geometrical and topological effects in physics.
In 1959 Aharonov and Bohm(AB)~\cite{aha} demonstrated that a quantum charge circulating  around a  magnetic flux tube acquires a 
quantum topological phase. This effect was observed
experimentally by Chambers\cite{prl:cham, pes1}. Aharonov and Casher showed that a particle with a magnetic moment moving 
in an electric field accumulates a quantum phase\cite{cas}, which has been
observed in a neutron interferometer\cite{cim} and in a neutral atomic Ramsey interferometer\cite{san}.
He and McKellar\cite{mac}, and Wilkens\cite{wil}, independently,
 predicted the existence of a quantum phase acquired by an electric
dipole, while circulating around, and parallel to a
line of magnetic monopoles. 
A  simple practical experimental configuration to test this phase,
was proposed by Wei, Han and Wei\cite{wei}. In their work the electric
field of a charged wire polarizes a neutral atom and when an uniform
magnetic field is applied parallel to the wire. In a recent article, a topological phase effect was proposed by Anandan\cite{ana} which describes an unified and fully
relativistic treatment of the interaction between a particle with permanent electric and magnetic dipole moments and an
 electromagnetic field.
The interaction of an electromagnetic field with a charged particle, plays a important role in the generation of colletive phenomena, for instance,  fractional statistics\cite{prange} and the quantum Hall effect.
The quantum motion of a charged particle in the presence of a constant magnetic field is described by Landau theory\cite{zlan}. The Landau quantization in two dimensions makes the energy levels coalesce into a discrete spectrum. The Landau levels  present a remarkable interest from many points of view. Its is the simplest model necessary for the description of the quantum Hall effect\cite{prange}. On the other hand, the Landau levels were studied for different two-dimensional surfaces\cite{ap1,ap2} with the interest in several areas of physics. Paredes {\it et al.}, using the analogy between a rotating Bose-Einstein condensate and a system of interacting electrons in a uniform magnetic field, proved the existence of  anyonic excitation in this condensates. Recently, Ericsson and Sj\"oqvist\cite{pra:sjo}, motivated by the results of Paredes {\it et al}\cite{prl:par,ssc:par}  proposed the first step towards another atomic Hall effect analogy. They  used  the Aharonov-Casher interaction, for a neutral particle with a permanent magnetic dipole, and proposed the analogy with the Landau levels quantization, for certain field-dipole configurations, that they have  denominated of Landau-Aharonov-Casher levels. Result is interesting and suggests the possibility of quantum Hall effect for magnetic dipoles in the presence of an electric field. Based on this idea we will analyze in this paper an analog of Landau quantization for a neutral particle that possesss a permanent  electric dipole. We will use the interaction of He-McKellar-Wilkens for an electric dipole in the presence of magnetic field. In the  same way as the Landau-Aharonov-Casher levels,  the particles can interact with the magnetic field  via a nonvanishing electric dipole. We use the Hamiltonian found by Anandan\cite{pla:ana} to describe the electric dipole in the presence of magnetic field. In this letter we adopt the systems of unity were $\hbar=c=1$.


Now, we briefly outline the  Ericsson and Sj\"oqvist~\cite{pra:sjo}  theory to describe the Landau-Aharonov-Casher(LAC) effect.  This theory  exhibits a similar Landau quantization for a neutral particle with a magnetic dipole  moving in an electric field. This quantization is demonstrated if  precise conditions in field-dipole configuration are obeyed. We use the Schr\"odinger equation approach to describe this theory. Our objective is to find explicitly the eingenfunctions and the degeneracy of energy levels. The choice of this approach to solve this problem is the fact that  we obtain explicit  wavefunctions  of the LAC levels.We adopt the following  cylindrical electric field configuration

\begin{equation}\label{eqc5.1}
	\mathbf{E}=\frac{\rho_{e} } {2} r \hat{e}_{r}\;,
\end{equation}
where $\rho_{e}$ is a nonvanishing uniform charge density. The Hamiltonian that describes  a neutral particle that possesses a nonvanishing magnetic moment  $\bm{\mu}$, in the presence of the electric field described by (\ref{eqc5.1}), in the nonrelativistic limit, is given by\cite{ana}
\begin{equation}\label{eqc5.2}
	H=\frac{1}{2M}(\mathbf{p}-\mu\mathbf{n}\times\mathbf{E})^2+\frac{\mu}{2M}\bm{\nabla}\cdot\mathbf{E}\;,
\end{equation}
where  $\mu$ is the intensity of the moment of magnetic dipole of the particle and $\mathbf{n}$ its direction. The Hamiltonian (\ref{eqc5.2}) presents an analogy to the minimal coupling for a charged particle in the presence of the magnetic field. We can define then the potential vector of Aharonov -- Casher as being 
\begin{eqnarray}
	\mathbf{A}_{AC}=\mathbf{n}\times\mathbf{E}
\end{eqnarray}
using the electric field given by (\ref{eqc5.1}) we obtain the following Aharonov-Casher potential
\begin{eqnarray}
	\mathbf{A}_{AC}=\frac{\rho_{e}}{2}r\hat{e}_{\phi}\;,\label{eqc5.3}
\end{eqnarray}
Using the definition (\ref{eqc5.3}) Ericsson and Sj\"oqvist defined the "magnetic" field associated with this vector potential. Choosing the dipole aligned parallel with the direction  $z$, $\mathbf{n}=\hat{e}_z$, we have the following Aharonov-Casher magnetic field
\begin{eqnarray}
	\mathbf{B}_{AC}&=&\bm{\nabla}\times\mathbf{A}_{AC}\nonumber\\
	&=&\rho_{e}\hat{e}_z\;.\label{eqc5.4}
\end{eqnarray}
Note that this configuration of Aharonov-Casher magnetic field is uniform. This configuration with the dipole configuration and movement of dipole restricted to the plane satisfies the conditions demonstred by Ericsson and Sj\"oqvist to obtain the analogous of Landau Levels, that are: $\mathbf{B}_{AC}$ uniform , absence of torque on the dipole and electrostatic conditions $\partial_{t}\mathbf{E}=0$ and $\bm\nabla \times\mathbf{E}=0$. All these conditions are satisfied by the configuration presented here that are similar to configurations presented in ref. \cite{pra:sjo}.

In this way we write the Schr\"odinger equation for this system,  in cylindrical coordinates, in the following form
\begin{eqnarray}\label{eqc5.5}
	-\frac{1}{2M}\left[\frac{1}{r}\frac{\partial}{\partial r}\left(r\frac{\partial\psi}{\partial r}\right)+\frac{1}{r^2}\frac{\partial^2\psi}{\partial\phi^2}\right]+ \\ \nonumber-\frac{i\omega}{2}\frac{\partial\psi}{\partial\phi}+\frac{M\omega^2}{8}r^2\psi+ \frac{\omega}{2}\psi={\cal{E}}\psi\;,
\end{eqnarray}
with the cyclotron frequency given by
\begin{equation}\label{eqc5.6}
	\omega=\sigma \omega_{AC}=\sigma \frac{|\mu\rho_{e}|}{M}.
\end{equation}
where $\sigma=\pm$.
 We use the following Ansatz to the solution of Scrhr\"odinger equation
\begin{equation}\label{eqc5.7}
	\psi=e^{i \ell \phi}R(r)\;,
\end{equation}
where  $\ell$ is an interger number. Using  Eq. (\ref{eqc5.7}),   Eq. (\ref{eqc5.5}) assumes the following form:
\begin{eqnarray}\label{eqc5.8}
	\frac{1}{2M}\left(R''+\frac1r R'-\frac{m^2}{r^2}R\right)+ \nonumber \\ + \left({\cal{E}}-\frac{M\omega_{AC}^2}{8}r^2+\frac{\sigma\ell\omega_{AC}}{2}- \frac{\sigma\omega_{AC}}{2}\right)R=0\;.
\end{eqnarray}
Now, We use the  following change of variables
\begin{eqnarray}
	\xi&=&\frac{M\omega_{AC}}{2}r^2.\;
\end{eqnarray}
In this way, Eq. (\ref{eqc5.8}) is transformed  into
\begin{equation}\label{eqc5.10}
	\xi R''+R'+\left(-\frac{\xi}{4}+\beta-\frac{\ell^2}{4\xi}\right)R=0\;,
\end{equation}
where we define 
\begin{equation}\label{eqc5.11}
	\beta=\frac{{\cal{E}}}{\omega_{AC}}+\frac{\sigma(\ell-1)}{2}\;.
\end{equation}
Studying the assymptotic limit of the solutions to Eq. (\ref{eqc5.10}) we can write the solution in the form
\begin{equation}\label{eqc5.12}
	R(\xi)=e^{-\xi/2}\xi^{|\ell|/2}\zeta(\xi)\;.
\end{equation}
We obtain a  hypergeometric equation that is satisfied by  the function  $\zeta(\xi)$   given by 
\begin{equation}\label{eqc5.13}
	\zeta=F\left[-\left(\beta-\frac{|\ell|+1}{2}\right),|\ell|+1,\xi\right]\;.
\end{equation}
Then, the energy is given by
\begin{eqnarray}
{\cal{E}}=\left(\nu +\dfrac{|\ell|}{2}-\frac{\sigma\ell}{2}+\frac{\sigma}{2}+\dfrac{1}{2}\right)\omega_{AC}\;,\label{eqc5.14}
\end{eqnarray}
where $\nu= 0, \pm 1, \pm2 \ldots$
The radial energy eingenfuctions is given by
\begin{eqnarray}\label{eqc5.15}
	R_{n,\ell}(r)&=&\frac{1}{a^{|\ell|+1}}\left[\frac{(|\ell|+\nu)!}{2^{|\ell|}\nu!|\ell|!^2}\right]\exp \left(-\frac{r^2}{4a^2}\right)\times \nonumber \\  &\times& r^{|\ell|}F\left[-\nu,|\ell|+1,\frac{r^2}{2a^2}\right]\;,
\end{eqnarray}
where
\begin{equation*}
a=a_{AC}=\sqrt{\frac{1}{M\omega_{AC}}}\;.
\end{equation*}
In a analogous way to the Landau levels, the energy levels are infinitely degenerated due to  magnetic translational symmetry. Here, the energy levels are infitely degenerated also. Ericsson and Sj\"oqvist  observed that the  energy levels are $\sigma$ dependent. In this way, the levels depend on the revolution axis direction. 
The eingenvalues are independent of the  orbit center, but are dependent of the revolution direction.

Now, we concentrate in the analysis of a Landau levels analogue  for the quantum dynamics of an electric dipole in the presence of an external magnetic field. The central idea of this letter is similar to the approach  developed by Ericsson and Sj\"oqvist to the Landau-Aharonov-Casher levels, presented previously in this letter, using a Schr\"odinger equation aproach. We use the He-McKellar-Wilkens interaction of the electric dipole to describe a new analogue of Landau levels for the electric dipole.  We consider a neutral particle that has a non null electric dipole moment $\mathbf{d}$. This particle is in moving in the presence of an external magnetic field $\mathbf{B}$. In our study of this problem we  demonstrate that in specific field-dipole configurations  we have a quantization similar to Landau levels. We consider a radial magnetic field in the following form
\begin{equation}\label{eq2.55}
	\mathbf{B}=\frac{\rho_{m}}{2}r\hat{e}_r\;.
\end{equation}
where $\rho_{m}$ is magnetic charge density. We can see clearly that this confguration is generated by a distribution of magnetic charge. The arrangement of field configurations with the magnetic field radially cylindrical  is more difficult to achieve experimentally, 
due to the fact that, {\it a priori}, we need a  distribution of magnetic charges. We can observe in the literature that this kind 
of arrangement would be possible. Some authors have claimed that this configuration can be obtained experimentally as in the 
arrangements presented  in the
articles\cite{mag,mag1,mag2}.  We choose the electric dipole to be  aligned in the z-direction. The  nonrelativistic Hamiltonian that describes the quantum dynamics of the  electric dipole, in the presence of the external field, is given by\cite{pla:ana}
\begin{equation}\label{eq2.56}
	H=\frac{1}{2M}(\mathbf{p}+d\mathbf{n}\times\mathbf{B})^2-\frac{d}{2M}\bm{\nabla}\cdot\mathbf{B}\;,
\end{equation}
where  $d$ is magnitude of electric dipole and  $\mathbf{n}$  is its direction. Now we define the vector potential of He-Mckellar-Wilkens,
$
	\mathbf{A}_{HMW}=\mathbf{n}\times\mathbf{B}
$. Using the field configuration adopted in Eq. (\ref{eq2.55}) we obtain the expression
\begin{eqnarray}
	\mathbf{A}_{HMW}=\frac{\rho_{m}}{2}r\mathbf{e}_\phi\;,\label{eq2.57}.
\end{eqnarray}
In this way, we obtain a "magnetic" field $\mathbf{B}_{HMW}=\bm{\nabla}\times(\mathbf{n}\times\mathbf{B})$ associated to the He-McKellar-Wilkens potential
\begin{eqnarray}
\mathbf{B}_{HMW}=\rho_{m}\hat{e}_z\;,\label{eq2.58}.
\end{eqnarray}
Note that  $d$ plays the role of a coupling constant\cite{pra:lee}. Here the same condition found by Ericsson and Sj\"oqvist\cite{pra:sjo} is obeyed for the existence of an analogue of Landau levels. The null torque condition  is guaranteed since  the speed of the particle is null in the direction
$\hat{e}_z$. Using  Eq.(\ref{eq2.58}) we have that $\mathbf{B}_{HMW}$ is homogeneous. It obeys the necessary conditions for the existence of a analogue of Landau Levels. The Schr\"odinger equation is therefore
\begin{eqnarray}\label{eq2.59}
&\dfrac{1}{2M}(\mathbf{p}+d\mathbf{A}_{HMW})^2\psi-\dfrac{d}{2M}\nabla\cdot\mathbf{B}\psi=
	{\cal{E}}\psi\;;&.
\end{eqnarray}	
Making use of Eqs. (\ref{eq2.55}), (\ref{eq2.57}) and (\ref{eq2.58}) into Eq.
(\ref{eq2.59}), in cylindrical coordinates, we have
\begin{eqnarray}
	-\dfrac{1}{2M}\left[\dfrac1r\dfrac{\partial}{\partial r}\left(r\dfrac{\partial\psi}{\partial r}\right)+\dfrac{1}{r^2}\dfrac{\partial^2\psi}{\partial\phi^2}\right]- \nonumber \\ -\dfrac{i\omega}{2}\dfrac{\partial\psi}{\partial\phi}+ \dfrac{M\omega^2}{8}r^2\psi-\dfrac{\omega}{2}\psi={\cal{E}}\psi\;,\label{eq2.60}
\end{eqnarray}
where we define the cyclotron frequency with
\begin{equation}\label{eq2.61}
	\omega=\sigma \omega_{HMW}=\frac{\sigma|d\rho_{m}|}{M},
\end{equation}
where $\sigma=\pm$. This equation is solved using the following Ansatz
\begin{equation}\label{eq2.62}
	\psi=Ce^{i\ell \phi}R(r)\;.
\end{equation}
where $\ell$ is an interger number and $C$ is a normalization constant. Substitut this wave function (\ref{eq2.62}) into the Schr\"odinger equation (\ref{eq2.60}) we obtain the radial equation
\begin{eqnarray}
	\dfrac{1}{2M}\left(R''+\dfrac{1}{r}{R'}-\dfrac{\ell^2}{r^2}R\right)+ \nonumber \\ +\left({\cal{E}}- \dfrac{M\omega_{HMW}^2}{8}r^2-\dfrac{\sigma\ell\omega_{HMW}}{2}+\dfrac{\sigma\omega_{HMW}}{2}\right)R=0\;.\label{eq2.63}.
\end{eqnarray}
Now, by using the change of variable $\xi=\frac{M\omega_{HMW}}{2}r^2\ $, Eq.
(\ref{eq2.63}) is transformed into
\begin{eqnarray}
	&\xi R''+R'+\left(-\dfrac{\xi}{4}+\beta-\dfrac{\ell^2}{4\xi}\right)R=0\;,&\label{eq2.65}
\end{eqnarray}
where
\begin{eqnarray}
	\beta=\frac{{\cal{E}}}{\omega_{HMW}}-\frac{\sigma(\ell-1)}{2}\;.\label{eq2.66}
\end{eqnarray}
Assuming, for the radial eigenfuction, the form
\begin{equation}\label{eq2.67}
	R(\xi)=e^{-\xi/2}\xi^{|\ell|/2}\zeta(\xi)\;.,
\end{equation}
which satisfies the usual asymptotic requirements and the finiteness at the origin for the bound state, we have
\begin{equation}\label{eq2.68}
	\xi\frac{d^{2}\zeta}{d\xi^{2}} +\left[(|\ell|+1)-\xi\right]\frac{d\zeta}{d\xi} -\gamma  \zeta=0
\end{equation}
where $\gamma=\beta-\frac{|\ell|+1}{2}$. We find that the solution of equation (\ref{eq2.68}) is the degenerated hypergeometric function
 \begin{eqnarray}
\label{hyper}
\zeta(\xi)=F\left[-\gamma,|\ell|+1,\xi\right]\;.
\end{eqnarray} 
In  order to have normalization of the wavefunction, the series in (\ref{hyper}) must be a polynomial of degree $\nu$, therefore
\begin{equation*}
	\gamma=\beta-\frac{|\ell|+1}{2}=\nu .
\end{equation*}
With this condition, we obtain discrete values for the energy, given by
\begin{equation}\label{eq2.69}
	{\cal{E}}_{\nu,\ell}=\left(\nu+\frac{|\ell|}{2}+\frac{\sigma\ell}{2}-\frac{\sigma}{2}+\frac12\right)\omega_{HMW}\;.
\end{equation}
The radial eigenfuction is then given by
\begin{eqnarray}\label{eq2.70}
	R_{\nu,\ell}&=&\frac{1}{a^{1+|\ell|}}\left[\frac{(|\ell|+\nu)!}{2^{|\ell|}\nu!|\ell|!^2}\right]^{1/2} \exp\left(-\frac{r^2}{4a^2}\right)\times \nonumber \\ &\times& r^{|\ell|}F\left[-\nu,|\ell|+1,\frac{r^2}{2a^2}\right]\;,	
\end{eqnarray}
where $a=(M\omega_{HMW})^{-1/2}$\;. Note that the energy levels are infinitely degenerated  similarly to the Landau levels. We observed a dependence of levels on the revolution direction.  $\sigma$ dependence on set of degenerate state define the LHMW problem, in analog way as the LAC levels but with a difference that of the revolution direction are opposed. Now we use ladder operator techiniques to demonstrate the relation  betwen the wave equation formalism adopted by us and the ladder operator adopted in \cite{pra:sjo}. We consider the Hamiltonian given by (\ref{eq2.56}) and the expression (\ref{eq2.57}) in Euclidean coordinate. The expression for the "vector" potential (\ref{eq2.57}) is similar to the symmetrical gauge of Landau levels, $\vec{A}_{HMW}=\frac{\rho_{m}}{2}(-y\hat{x}+x\hat{y})$ ,and we choose the following operators to describe the problem

\begin{eqnarray}\label{ope}
\hat{a}&=&\frac{1}{2M|\omega_{HMW}|}(\Pi_{x} + i\sigma\Pi_{y})\\ \nonumber
\hat{a}^{\dagger}&=&\frac{1}{2M|\omega_{HMW}|}(\Pi_{x} - i\sigma\Pi_{y}),
\end{eqnarray}
where $\Pi=(-\nabla +d\vec{n}\times \vec{B})$ and with the commutation relation $[\hat{a},\hat{a}^{\dagger}]=1$. Using (\ref{ope}) and the Hamiltonian (\ref{eq2.56}) we obtain
\begin{eqnarray}
H=[\hat{a}^{\dagger}\hat{a} +\frac{1}{2}(1-\sigma)]\omega_{HMW}.
\end{eqnarray} 
 In this way, we obtain the following energy eigenvalues
\begin{eqnarray}\label{eneope}
E_{n,\sigma,l}=[n +\frac{1}{2}(1-\sigma)]\omega_{HMW}.
\end{eqnarray} 
Note that  the expression (\ref{eneope}) can be related to the expression (\ref{eq2.69}) by the indentification $n=\nu+\frac{|\ell|}{2}+\frac{\sigma\ell}{2}$. 

In this letter  we study the energy eingefunctions and the eigenvalues  of a neutral particle with a permanent electric dipole moment in the presence an  external magnetic field. We  demonstrated that the HMW interaction is responsible for  the quantization of energy of the particle in a similar way of the AC interactions in the LAC levels investigated by Ericsson and Sj\"oqvist\cite{pra:sjo}. We have used the Hamiltonian found  by Anandan to describe this system and solved the related Schr\"odinger equation. It is interesting to observe that, in the  same way of the LAC levels, the LHMC levels are also dependent on the revolution direction but have  an  opposite direction. This result can be explained if we consider the duality between these two problems. The equations of motion for LHMW has the same form as the equation  for LAC. Indeed, changing $-d$ by $\mu$ and $\mathbf{B}$ by $\mathbf{E}$ we obtain the equation of motion for the latter. The physical quantities, in the effect, we had consider in  our work and the LAC levels, studied in~\cite{pra:sjo}, are  related by a dual rotation. In this sense the LHMW levels are the dual of  the LAC levels. We emphasize that when we perform  the duality transformation the dipole  revolution direction is also modified. With advances in cold atom\cite{prl:kuk,prl:duan} technology it is possible to simulate this effect studied in this letter. The cold Rydberg atoms have been explored as  systems for a possible test of the LHMW levels in  more realistic magnetic field configurations. We call attention that more realistic configurations of magnetic field  can generate similar effects, as for example the Wei and Han\cite{wei} field-dipole configurations, which can be used to study a possibility of  analog Landau quantization due to an induced electric dipole in cold atom systems. This subject will be presented in a future publication.
\acknowledgments
This work was partially supported by CNPq, CAPES/PROCAD, CNPQ/FINEP/PADCT and PRONEX/CNPQ/FAPESQ .We thank Professor F. Moraes for the critical reading of this manuscript.

\end{document}